\documentclass[twocolumn,preprintnumbers]{revtex4}
\usepackage{graphicx,epsfig}

\begin{document}

\preprint{CLNS~04/1891, LMU~15/04, PITHA~04/17, SHEP~04/37, hep-ph/0411171}

\title{\boldmath 
Comment on ``$B\to M_1 M_2$: Factorization, charming penguins,\\ 
strong phases, and polarization''\unboldmath}

\author{M.~Beneke${}^1$, G.~Buchalla${}^2$, M.~Neubert${}^{3,4}$, and 
C.T.~Sachrajda${}^5$}

\affiliation{
$^1$Institut f\"ur Theoretische Physik E, RWTH Aachen, D--52056 Aachen, 
Germany\\
$^2$Ludwig-Maximilians-Universit\"at M\"unchen, Sektion Physik, D--80333 
M\"unchen, Germany\\
$^3$Laboratory for Elementary-Particle Physics, Cornell University, Ithaca, 
NY 14853, U.S.A.\\
$^4$School of Natural Sciences, Institute for Advanced Study, Princeton, 
NJ 08540, U.S.A.\\
$^5$School of Physics and Astronomy, University of Southampton, 
Southampton SO17 1BJ, U.K.}

\date{September 19, 2004}

\begin{abstract}\noindent
We show that the factorization formula for non-leptonic $B$ decays to two 
light flavor non-singlet mesons derived by Bauer et al.\ in the context of 
soft-collinear effective theory is equivalent to the corresponding formula in 
the QCD factorization approach. The apparent numerical differences in the 
analysis of $B\to\pi\pi$ data performed by these authors, as compared to 
previous QCD factorization analyses, can largely be attributed to the neglect 
of known perturbative and power corrections. 
\end{abstract}

\maketitle

The extent to which hadronic decays of $B$ mesons to two light hadrons can be 
computed from first principles in QCD has been the subject of many 
investigations, following the statement \cite{Beneke:1999br,Beneke:2000ry} 
that the decay amplitudes factorize in the limit of very large $B$-meson mass. 
Taking the final state to consist of two pions, the amplitude can be 
represented schematically in the form 
\begin{equation}\label{ff}
   A(B\to \pi\pi) = F^{B\to\pi}\,T^{\rm I}\star\Phi_\pi
   + T^{\rm II}\star\Phi_B\star\Phi_\pi\star\Phi_\pi \,.
\end{equation}
In this equation, $F^{B\to\pi}$ denotes a physical form factor, and $\Phi_B$ 
and $\Phi_\pi$ are the leading-twist light-cone distribution amplitudes of the 
mesons. The quantities $T^{\rm I,II}$ are perturbative hard-scattering kernels 
involving the two scales $m_b$ (hard) and $\sqrt{m_b\Lambda}$ 
(hard-collinear), which are linked to the other elements of the formula by 
convolution integrals (indicated by an asterisk).

The authors of \cite{Bauer:2004tj} suggest a factorization formula 
(their Eq.~(24)) similar to 
(\ref{ff}) in the framework of soft-collinear effective theory (SCET) and 
imply that it is conceptually different. They also argue that, even at leading 
order in the $1/m_b$ expansion, there may be an additional term on the 
right-hand 
side of (\ref{ff}), corresponding to long-distance contributions from 
$c\bar c$ penguins. They do not disprove factorization of charm-penguin loops 
by providing a counter-example to factorization; rather, they state that they 
were not able to demonstrate factorization.

In this Comment, we explain why the formula given in \cite{Bauer:2004tj} is 
identical in content to the QCD factorization formula (\ref{ff}), and why 
non-factorizable charm-penguin contributions are of higher order in the 
$1/m_b$ expansion. We also point out that the phenomenological analysis of 
\cite{Bauer:2004tj} neglects important perturbative and power corrections 
which are already known. Once these are included, there is little room for 
significant additional contributions to the QCD penguin amplitude.

\section{Equivalence of the SCET and QCD factorization formulae}

We first note that the coefficient function of the spectator-scattering term 
can be represented as a convolution $T^{\rm II}=C^{\rm II}\star J$ of hard and 
hard-collinear coefficient functions. The formula quoted in 
\cite{Bauer:2004tj} follows from (\ref{ff}) by rewriting 
\begin{equation}\label{rearrange}
   T^{\rm II}\star\Phi_B\star\Phi_\pi\star\Phi_\pi
   = C^{\rm II}\star\zeta_J^{B\pi}\star\Phi_\pi \,,
\end{equation}
with $\zeta_J^{B\pi}$ defined as $J\star\Phi_B\star\Phi_\pi$, and $C^{\rm II}$ 
defined by the decomposition of $T^{\rm II}$ shown above. In addition, in 
\cite{Bauer:2004tj} the SCET form factor $\zeta^{B\pi}$ rather than the 
physical QCD form factor $F^{B\to\pi}$ is used. As discussed in 
\cite{Beneke:2000ry}, this implies another rearrangement of this type.

In general SCET provides a powerful tool to simplify
factorization proofs (a task that has not yet been completed for 
the case of $B\to\pi\pi$ considered here), but the resulting QCD 
factorization formulae can also be obtained using traditional factorization 
methods, as is frequently done in practice.

The authors of \cite{Bauer:2004tj} entertain the possibility that the 
hard-collinear scale may be non-perturbative, and hence they choose not to 
factorize $\zeta_J^{B\pi}$ into $J\star\Phi_B\star\Phi_\pi$ as indicated 
above. This is a logical possibility in the QCD factorization approach. 
However, we show below that it is not supported by theoretical calculations. 

We disagree with \cite{Bauer:2004tj} on the statement that 
$\zeta_J^{B\pi}\ll F^{B\to\pi}$ is a prediction of QCD factorization, whereas 
the SCET treatment suggests $\zeta_J^{B\pi}\sim F^{B\to\pi}$. The 
factorization formula states that both terms in (\ref{ff}) are of the same 
order in $1/m_b$ power counting and that $T^{\rm I}$ and $T^{\rm II}$ start at 
$O(\alpha_s^0)$ and $O(\alpha_s(\sqrt{m_b\Lambda}))$, respectively. However, 
it does not predict the relative size of the two terms, and as we explain 
below, the numerical value of $\zeta_J^{B\pi}$ depends on several 
hadronic input parameters, which are rather uncertain at present.

The authors of \cite{Bauer:2004tj} point out that the hard-collinear kernel 
$J$ is universal, so that only a single function $\zeta_J^{B\pi}$ appears in 
(\ref{rearrange}), which is the same that appears in the factorization of the 
$B\to\pi$ form factors. This fact is important for phenomenology when one opts 
to treat the hard-collinear scale as non-perturbative, but does not by itself 
represent a conceptual difference between the formulae given in 
\cite{Beneke:1999br,Beneke:2000ry} and \cite{Bauer:2004tj}. Furthermore, the 
usefulness of the universality of $J$ is limited to the approximation where 
one neglects radiative corrections to the hard-scattering kernels 
$C^{\rm II}$, since only then does the function $\zeta_J^{B\pi}$ reduce to a 
single number \cite{Beneke:2003pa,Hill:2004if}. In other words, a 
phenomenological treatment of the hard-collinear scale as non-perturbative 
relies on the approximation that the kernels are restricted to their 
tree-level approximations, whereas one of the key features of QCD 
factorization (as opposed to naive factorization) is that one can consistently 
include radiative corrections. 

\section{Charm-penguin loops}

\begin{figure}
\begin{center}
\epsfig{file=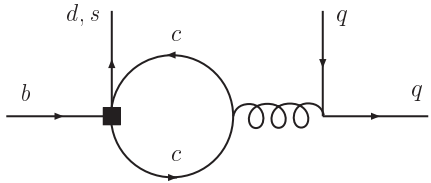,width=6.0cm}
\end{center}
\vspace{-0.4cm}
\caption{\label{fig:penguin}
Charm-penguin loop.}
\end{figure}

In \cite{Bauer:2004tj} it is claimed that there is a possible exception to 
factorization from diagrams with charm-quark loops (see 
Figure~\ref{fig:penguin}). The argument is based on the observation that when 
the gluon virtuality is near the $c\bar c$ threshold, $q^2\approx 4 m_c^2$, 
the non-relativistic scales $m_c v$ and $m_c v^2$ become important. Since 
$m_c v^2\approx\Lambda$ numerically, this appears to introduce a sensitivity 
to non-perturbative scales without power-suppression in $1/m_b$. The authors 
of \cite{Bauer:2004tj} suggest that these diagrams do not factorize, in the 
sense that the long-distance physics at leading-power in the $1/m_b$ expansion 
cannot be factored into form factors or light-cone distribution amplitudes. We 
emphasize that the question of non-factorizable $c\bar c$ effects at leading 
order in the heavy-quark expansion is a different issue than the one raised in 
\cite{Ciuchini:2001gv}, where it is speculated that power corrections to the 
QCD charm-penguin amplitudes may be numerically large. 

Factorization statements, be they derived 
diagrammatically or with soft-collinear effective theory, always concern 
properties of the amplitude in certain asymptotic limits, here an expansion in 
$1/m_b$, independent of the actual size of the expansion parameter. It is 
therefore important to clearly distinguish the issue of factorization at 
leading order in the $1/m_b$ expansion from the question of whether there are 
non-factorizable contributions which are formally power-suppressed, but which 
nevertheless may be numerically significant. In the following discussion, we 
focus on the question of factorization in the formal heavy-quark limit.

The most intuitive way of understanding why the threshold region does not 
require special treatment is based on quark-hadron duality 
\cite{Beneke:2000ry}. The integration over the gluon virtuality in the range 
$0\le q^2\le m_b^2$ is weighted by the pion distribution amplitude, which is 
smooth over the entire integration region. This provides the necessary 
smearing of the loop amplitude, which ensures that the result is given by a 
simple partonic calculation up to power corrections, in complete analogy with 
the standard justification of the partonic interpretation of inclusive 
heavy-meson decays, or cross sections in general. The smearing also ensures 
that one can apply the same power-counting arguments that demonstrate 
factorization of other diagrams to charm-penguin diagrams with no need to 
single out the threshold region. In particular, in the non-relativistic 
situation implied in \cite{Bauer:2004tj} there is no need to sum Coulomb 
ladder diagrams, since these do not result in large perturbative corrections 
after the integration over $q^2$. 
 
Even without invoking the duality argument, the fact that the charm threshold 
region comprises only a parametrically small portion of the entire integration 
implies a phase-space suppression. This fact has been neglected in the 
argument of \cite{Bauer:2004tj}. More precisely, writing $q^2=\bar x m_b^2$, 
where $\bar x$ denotes the longitudinal momentum fraction of the anti-quark in 
one of the pions, this region is $\Delta\bar x\sim v^2\,(m_c/m_b)^2$ or 
$\Delta\bar x\sim\Lambda m_c/m_b^2$, whichever is larger. In order to study 
the question of whether long-distance $c\bar c$ loop effects are of leading 
order or not, it is necessary to decide how the limit $m_b\to\infty$ is to be 
taken. If we define the heavy-quark limit by $m_b\to\infty$ with $m_c$ fixed, 
one may distinguish several possibilities such as $m_c\sim\Lambda$, 
$m_c v^2\sim\Lambda$, or $m_c v^2\gg\Lambda$. While the physics of the 
threshold region is very different for all these cases, they share the common 
feature that, for the purposes of power counting, the charm quark can be 
considered to be a light quark, and the suppression of long-distance 
$c\bar{c}$ effects has the same origin as that for the corresponding diagrams 
with light-quark loops, which implies $\Delta\bar x\sim 1/m_b^2$. In addition, 
since in this region $\bar x\sim m_c^2/m_b^2$, there is a further suppression 
due to the end-point behavior of the pion distribution amplitude (which 
vanishes linearly as $\bar x\to 0$).

If we define the heavy-quark limit as $m_{b,c}\to\infty$ with the ratio 
$m_c/m_b$ fixed, then there is no power suppression due to the phase-space or 
end-point behavior of the distribution amplitudes, but the threshold region is 
perturbative, up to a small non-perturbative contribution of order 
$v^2\cdot v^2\,(\Lambda/(m_c v^2))^4\sim(\Lambda/m_b)^4$ 
\cite{Voloshin:1979uv}. 

We conclude that charm-penguin diagrams factorize at leading power in $1/m_b$. 
The argument for factorization remains valid also for more complicated 
higher-order penguin graphs whenever the threshold region is phase-space (and 
end-point) suppressed or the charm quark is heavy so that perturbation theory 
is applicable.

\section{Validity of perturbation theory at the hard-collinear scale}

In applying the QCD factorization formula to phenomenology the authors of 
\cite{Bauer:2004tj} treat the hard-collinear scale $\sqrt{m_b\Lambda}$ as 
non-perturbative, and hence the quantity $\zeta_J^{B\pi}$ as an unknown 
phenomenological function. This is justified {\em a posteriori\/} following 
the result of a phenomenological fit (on which we comment below). This line of 
argument ignores the fact that perturbation theory at scales of order 
$\sqrt{m_b\Lambda}\sim m_c$ has been used successfully in many important 
applications in $B$-physics, including all determinations of $|V_{ub}|$ from 
inclusive $B$-decays, and studies of the hadronic decay rate of the $\tau$ 
lepton. As already mentioned, a serious drawback of treating the 
hard-collinear scale as non-perturbative is that it renders the factorization 
approach unpredictive beyond the tree approximation, because only the integral 
over $\zeta_J^{B\pi}$ and not its functional form can be extracted from 
measurements.

A systematic way to address the question of the perturbativity of the 
hard-collinear scale is to calculate higher-order corrections in $\alpha_s$ to 
the jet function $J$ in the product $T^{\rm II}=C^{\rm II}\star J$, defined as 
the Wilson coefficient function arising in the matching of certain (type-B) 
SCET$_{\rm I}$ 
current operators onto four-quark operators of SCET$_{\rm II}$. The 
next-to-leading order terms have been computed recently and were found to be 
small \cite{Hill:2004if}. Specifically, for the case of light pseudoscalar 
mesons and an asymptotic light-cone distribution amplitude 
$\Phi_\pi(x)=6x(1-x)$, one finds that the convolution integrals over the jet 
function give rise to the series
\begin{equation}\label{nlocorrection}
   \frac{\alpha_s(\mu_i)}{\lambda_B}
   \left[ 1 + {\alpha_s(\mu_i)\over\pi} \left( \frac{\langle L^2\rangle}{3}
   - 1.31 \langle L\rangle + 1.00 \right) + \dots \right] ,
\end{equation}
where $\mu_i\sim\sqrt{m_b\Lambda}$ is the hard-collinear scale, $\lambda_B$ is 
the first inverse moment of the $B$-meson distribution amplitude 
$\Phi_B(\omega,\mu_i)$, $L=\ln(m_b\omega/\mu_i^2)$, and $\langle\dots\rangle$ 
denotes an average over $\Phi_B(\omega,\mu_i)$ with measure $d\omega/\omega$. 
While the precise form of  the $B$-meson distribution amplitude is unknown, 
the fact that $\omega\sim\Lambda$ ensures that $L$ cannot be large, giving a 
small coefficient to the next-to-leading term. For example, using the results 
of \cite{Braun:2003wx} for the moments $\langle L^2\rangle$, 
$\langle L\rangle$ the coefficient of $\alpha_s/\pi$ in (\ref{nlocorrection}) 
is $2.2\pm 0.6$ for $\mu_i^2=0.5\,\mbox{GeV}\,m_b$. There is thus no evidence 
that perturbation theory cannot be applied at the hard-collinear scale. We 
also note in this context that the power corrections from the hard-collinear 
scale are $1/m_b$ suppressed (and not $1/\sqrt{m_b}$) just as those from the 
hard scale.

Since the perturbative corrections to the jet function are well behaved, the 
quantity $\zeta_J^{B\pi}$ can be factorized and expressed in terms of 
convolution integrals over light-cone distribution amplitudes. The question of 
the numerical value of $\zeta_J^{B\pi}$, and whether it is a small 
contribution to the physical form factor $F^{B\to\pi}$, rests on the 
properties of these amplitudes, as well as on other parameters such as the 
strange-quark mass. At leading order in perturbation theory, we obtain for 
$\zeta_J^{B\pi}$ the result 
\begin{equation}
   \zeta_J^{B\pi} = \frac{3\pi\alpha_s C_F}{N_c^2}\,
   \frac{f_B f_\pi}{M_B\lambda_B}\left(\langle \bar y^{-1}\rangle_\pi
   + r_\chi^\pi X_H\right) 
\end{equation}
where the notations of \cite{Beneke:2003zv} have been used. Taking the default 
values and uncertainties of the input parameters from this reference, and 
adding errors in quadrature, yields $\zeta_J^{B\pi}=0.016\mbox{--}0.064$, 
which is small compared with typical values $F^{B\to\pi}=0.24\mbox{--}0.30$. 
($F^{B\to\pi}=\zeta^{B\pi}+\zeta_J^{B\pi}$ when hard matching corrections are 
neglected.) Taking some correlated parameter variations so as to reproduce the 
data on $B\to\pi\pi$ decays, scenario S2 of \cite{Beneke:2003zv} yields the 
somewhat increased value $\zeta_J^{B\pi}=0.080$. To obtain significantly 
larger results would require a very small value of the hadronic parameter 
$\lambda_B$. While this is a logical possibility, a recent QCD sum-rule 
calculation of $\lambda_B$ gives a value around $0.45\,$GeV 
\cite{Braun:2003wx}, which is in fact somewhat larger than the estimate 
adopted in \cite{Beneke:1999br,Beneke:2000ry}. 

These estimates are to be compared to the fit result 
$\zeta_J^{B\pi}=0.11\pm 0.03$ obtained by the authors of \cite{Bauer:2004tj}, 
who also find that the bulk of the $B\to\pi$ form factor comes from 
$\zeta_J^{B\pi}$, giving the very small result $F^{B\to\pi}=0.17\pm 0.02$. 
This picture contradicts the QCD sum rules for heavy-to-light form factors, in 
which $\zeta_J^{B\pi}$ must be associated with a radiative correction 
\cite{Bagan:1997bp}. The preference of the data for a smaller $B\to\pi$ form 
factor together with an increase of the hard-spectator scattering contribution 
to the color-suppressed tree amplitude $a_2$ has already been discussed in 
\cite{Beneke:2003zv}, which however did not arrive at a similarly extreme 
conclusion. This discrepancy can be traced to a few omissions in the 
calculation of \cite{Bauer:2004tj}, each of which has a minor effect: the 
absence of radiative corrections, the absence of phases in tree amplitudes, 
the absence of the scalar up-penguin amplitude in $T_c$, the use of asymptotic 
wave functions, and finally, a larger value of $|V_{ub}|$. When these effects 
are taken into account and combined with the most recent experimental data, 
one finds a significantly smaller value of $\zeta_J^{B\pi}$ and a larger value 
of $F^{B\to\pi}$, in qualitative agreement with theoretical expectations.

\section{The QCD penguin amplitude}

We now turn to the discussion of the phenomenological analysis of the 
$B\to\pi\pi$ data performed in \cite{Bauer:2004tj}. Our principal criticism in 
addition to what has already been described concerns the evaluation of the QCD 
penguin amplitude. It may be written as 
\begin{equation}
   P = a_4 + r_\chi^\pi a_6 + \beta_3 \approx -0.09 \,,
\end{equation}
where $a_4\approx -0.023\,[\alpha_s^0]-0.002\,[\alpha_s^1]$ represents the 
vector penguin contribution, 
$r_\chi^\pi a_6\approx -0.038\,[\alpha_s^0]-0.014\,[\alpha_s^1]$ the 
$1/m_b$ suppressed scalar penguin contribution, and $\beta_3\approx -0.011$ a 
power suppressed and rather uncertain penguin annihilation term. (The numbers 
are based on the analysis in \cite{Beneke:2003zv}. Without errors they should 
be taken only for illustration purposes. In particular, we neglected all 
phases, since they are unimportant for the following discussion.) 

In the calculation of \cite{Bauer:2004tj} the lowest order ($\alpha_s^0$) and 
some of the $\alpha_s$ contributions to $a_4$ (those included in the 
phenomenological parameters for hard scattering and charm penguins) are taken 
into account. The term $r_\chi a_6+\beta_3$ is dropped, because it is 
power-suppressed. Now while it is true that the factorization properties of 
power-suppressed contributions in general, and scalar penguin contributions in 
particular, have not yet been investigated to all orders in perturbation 
theory, the large tree-level contribution to $r_\chi a_6$ suggests that if one 
neglects power corrections entirely (as done in \cite{Bauer:2004tj}) one is 
certain to obtain a poor approximation. We emphasize that this is unrelated to 
the charm-quark loops discussed above, which appear only in the small 
$\alpha_s$ corrections. By dropping the scalar penguin amplitude, the authors 
of \cite{Bauer:2004tj} are forced to erroneously assign the QCD penguin 
amplitude almost entirely to the charm-quark loops.

There is considerable phenomenological evidence that the scalar penguin 
amplitude is in approximate agreement with our theoretical expectations. The 
suppression of the pseudoscalar-vector and vector-pseudoscalar penguin 
amplitudes relative to the pseudoscalar-pseudo\-scalar penguin amplitude 
\cite{Beneke:2003zv}, as well as the pattern of drastically different 
branching fractions for the decay modes $B\to \eta^{(\prime)} K^{(*)}$ 
\cite{Beneke:2002jn}, can be attributed directly to the different size and 
sign of the $r_\chi a_6$ term relative to $a_4$ in the QCD penguin amplitude. 
We are unaware of any other theoretical framework that can explain these 
facts. From such studies of penguin dominated $B$-decays we are therefore led 
to the conclusion that there is little room for extra contributions to the QCD 
penguin amplitude.

\begin{acknowledgments}
We are grateful to Iain Stewart for useful discussions. M.B.\ and M.N.\ would
like to thank the KITP, Santa Barbara, for hospitality during the preparation 
of this note. The work of M.B.\ is supported in part by the DFG 
Sonderforschungsbereich/Transregio 9 ``Computer-gest\"utzte Theoretische 
Teilchenphysik''. The research of M.N.\ is supported by the National Science 
Foundation under Grant PHY-0355005, and by the Department of Energy under 
Grant DE-FG02-90ER40542. The work of C.T.S.\ is supported by PPARC Grant 
PPA/G/0/2002/00468.
\end{acknowledgments}

\end{document}